\documentclass[preprint]{aastex}

\newcommand{\cm}{{\mathrm{cm}}}
\newcommand{\he}{{\mathrm{He}}}
\newcommand{\BH}{{\mathrm{BH}}}
\newcommand{\Msun}{M_{\sun}}
\newcommand{\Mss}{\Msun\,{\mathrm{s}}^{-1}}
\newcommand{\Msyr}{\Msun\,{\mathrm{yr}}^{-1}}
\newcommand{\erg}{{\mathrm{ergs}}}
\newcommand{\es}{\erg\,s^{-1}}

\shortauthors{Zhang \& Fryer}
\shorttitle{THE MERGER OF A HELIUM STAR AND A BLACK HOLE}

\begin{document}

\title{THE MERGER OF A HELIUM STAR AND A BLACK HOLE: GAMMA-RAY BURSTS}
\author{Weiqun Zhang and Chris L. Fryer}
\affil{UCO/Lick Observatory, University of California
\\  Santa Cruz, CA 95064}
\email{zhang@ucolick.org and cfryer@ucolick.org}

\begin{abstract}

There is growing observational evidence that gamma-ray bursts (GRBs) 
are powered by black holes accreting rapidly through a disk.  The 
supernova-like outburst that accompanies some gamma-ray bursts 
suggest that {\it some} long-duration GRBs may be driven by the 
accretion of a rotating stellar core onto a central black hole.  
Such a system can be produced when a compact remnant spirals into 
the core of its binary companion.  During the inspiral, orbital 
angular momentum is injected into the helium core.  By the time 
the compact remnant reaches the center of the helium core, it too 
has gained angular momentum as well as mass, producing a rapidly 
accreting black hole (or neutron star) at the center of a rotating 
stellar core.  

Whether or not such a merger (termed He-merger) can produce a GRB 
depends upon the initial mass and spin of the central black hole, 
as well as the angular momentum in the stellar core.  In this paper, 
we use a 3-dimensional smooth particle hydrodynamics (SPH) code to 
follow the He-merger process and make quantitative estimates of the 
initial mass and spin of the central compact remnant, as well as the 
angular momentum in the accreting helium core.  During the inspiral, 
a $2\,\Msun$ compact remnant gains $\sim 0.5-3.5\,\Msun$ depending on the
mass of the helium core (more massive cores provide more accretion).
The accretion rates on the central remnant are initially very high,
$10^5\,\Msyr$ up to $10^6\,\Msyr$ (the accretion rate increases with
increasing helium star mass), and the central remnant quickly becomes a 
black hole if it was not one already.  

From these accretion rates, we estimate GRB explosion energies.  
In all mergers, magnetically driven jets are expected to  produce GRB
explosions with energies above $10^{51}$\,ergs.  
For neutrino-annihilation-driven 
explosions, the GRB energy increases dramatically with helium star mass:  
the merger of a $2\,\Msun$ compact remnant with a $4\,\Msun$ helium 
star only produces a $10^{47}\,$erg explosion in $\sim 500$\,s whereas 
the merger of a $2\,\Msun$ compact remnant with a $16\,\Msun$ helium 
star produces a $>10^{52}\,$erg explosion in $\sim 65$\,s.  

\end{abstract}
        

\section{Introduction} \label{sec:intro}

Although there is no agreed upon engine powering gamma-ray bursts(GRBs), 
models based upon black hole accretion disks (BHAD models) are 
attractive.  The progenitors of this class of 
GRB engine include the mergers of compact binaries (double neutron 
star, black hole and a neutron star, white dwarf and a black hole), 
collapsars, and the inspiral of a neutron star or black hole into a 
helium star (see Fryer, Woosley, \& Hartmann 1999a for a review).  
In all of these cases, these progenitors produce accretion 
disks around black holes which, as they accrete, use neutrinos 
or magnetic fields to convert part of the gravitational potential or 
rotational energy of the disk material (or black hole) into a 
GRB outburst.  The BHAD systems formed by these progenitors 
differ mainly in the disk mass, disk angular momentum, or amount 
of material above the plane of the rotation disk and these 
variations may well explain the diversity of the observed sample 
of gamma-ray bursts.

For instance, classical GRBs seem to divide into long and short
duration subgroups 
(e.g. Kouveliotou et al. 1993; Katz \& Canel 1996; Tavani 1998), 
and it is thought that these two distinct subgroups are explained 
by different BHAD progenitors (e.g. Fryer et al. 1999a).  Three 
progenitors may produce long-duration GRBs:  mergers of 
white dwarfs with black holes, collapsars, and He-mergers.  
The disks formed in the mergers of white dwarfs with 
black holes (or neutron stars\footnote{Belczy\'{n}ski, Bulik, \& 
Rudak (2000) have found that white dwarf/neutron star 
mergers, which may be able to produce GRB outbursts using a 
magnetic-field driven mechanism, dominate the number of white 
dwarf mergers.}) represent one extreme in disk characteristics.  
Of the three long-duration BHAD GRB progenitors, white dwarf mergers 
produce disks with the most angular momentum and the least amount 
of material along the rotation axis.  Although the high angular momentum 
in these disks make neutrino-mediated explosions extremely weak, 
white dwarf/black hole mergers may produce strong GRB explosions via some 
magnetic field mechanism (Fryer et al. 1999b).  Collapsars, on the 
other hand, represent the other extreme:  they have less angular 
momentum and much more material along the axis of rotation 
(MacFadyen \& Woosley 1999; MacFadyen, Woosley ,\& Heger 2000).  
Since the connection between GRB 980425 and supernova 1998bw 
(Kulkarni et al. 1998), which is naturally explained by collapsar 
models (Woosley 1993; MacFadyen \& Woosley 1999; MacFadyen et al. 2000),
the collapsar model has become the favorite 
progenitor for long-duration bursts.

But can all long-duration gamma-ray bursts be collapsars?
There is increasing evidence to suggest that collapsars cannot 
explain all GRBs.  First, the  X-ray and optical lightcurves 
of GRBs suggest that the environment around the GRB engine 
is sometimes best fit by a wind-swept model, and other times 
is best fit by a constant density model (Chevalier \& Li 2000, 
Livio \& Waxman 2000).  Based on the fact that collapsars occur 
only in massive stars, Chevalier \& Li (2000) and Livio \& Waxman 
(2000) both argue that those GRBs best fit by constant density 
environments must not be collapsars\footnote{However, bear in mind that 
collapsars form only with stars that had very low mass-loss rates
(presumably low metallicities) and the winds of collapsar progenitors 
may not alter the environment significantly (Fryer et al. 1999a).  
Even those GRBs which are best matched by constant density environments 
may yet be collapsars.}.  In addition, collapsars may have trouble 
explaining the iron emission lines detected in three GRBs:  GRB 970508 
(Piro et al. 1999), GRB 970828 (Yoshida et al. 1999), and GRB 991216 
(Piro, et al. 2000).  To explain these detections (the detections in 
GRB 970508 and GRB 970828 are marginal at best so not all may be real), 
B\"ottcher (2000) found that the progenitors of these bursts may need to 
have metallicities in excess of solar.  
However, the winds of collapsar progenitors must be lower 
than typical winds of massive stars, and hence collapsar progenitors 
are likely to be low-metallicity and it is unlikely that collapsars 
can explain such high iron abundances.

To compare the He-merger model to that of the collapsar, 
let us review the formation scenario of He-merger GRBs.  The 
progenitor of this GRB engine is a binary system consisting of 
two massive stars.  The more massive star (primary) collapses to form 
a neutron star or black hole.  The loss of matter in the 
supernova explosion and a possible kick on the compact remnant 
unbinds many of the binaries.  Those that remain bound tend 
to be in tight orbits, and when the companion star evolves 
off the main sequence, it engulfs the compact remnant and 
the system goes into a common envelope phase.  Once in 
a common envelope, the compact remnant spirals down 
through the hydrogen star.  If the binding energy of the 
envelope is too large, the compact star will spiral into 
the companion's helium core. It is this inspiral which spins up the 
helium core and produces the conditions necessary to form a gamma-ray 
burst.

The merger of the compact remnant with its helium companion provides 
a mechanism to produce BHAD system very similar to that of a 
collapsar (Fryer \& Woosley 1998).  The primary difficulty 
with the collapsar model is producing a black hole with a 
rotating stellar envelope.  The helium merger model naturally 
forms an accreting black hole surrounded by a {\it rotating} 
stellar core.  The differences in the formation processes 
of these two GRB models produce distinguishing features which 
can be observed.

For example, after the formation of the compact remnant, the 
He-merger binary system will have some net velocity with respect 
to its formation region, and it may travel beyond the wind and 
supernova ejecta of the primary.  Hence, the constant density 
environment required by some of the lightcurves is easily explained 
by He-mergers.  In addition, the primary supernova's explosion may 
enrich the hydrogen envelope of its companion (Israelian et al. 1999).  
When this envelope is ejected during the common envelope phase, 
it produces an iron rich torus which may explain the observed 
iron lines (B\"ottcher \& Fryer 2000).  

Fryer \& Woosley (1998) argued that the engine driving 
He-merger GRBs would be very similar to collapsars.  But 
to truly understand the He-merger engine, we must follow 
the inspiral and obtain quantitative estimates of both 
the rotation of the helium core and the remnant mass 
in the core.  Knowing these values, we can then infer 
the duration and total energy of the He-merger GRB.  
In this paper, we present hydrodynamical simulations of 
the inspiral of a compact remnant into a helium core for 
a range of helium core masses.  We describe our numerical 
techniques in \S~\ref{sec:nm} and the results of our simulations in 
\S~\ref{sec:res}.  We conclude with a discussion of the inspiral 
results and of the viability of He-mergers as a GRB model.

\section{Numerical Methods} \label{sec:nm}

He-merger GRBs are formed when, in a binary system, a giant 
star envelops its compact remnant (neutron star or black hole) 
companion, causing it to spiral through the hydrogen envelope 
in what is termed a ``common-envelope'' phase.  The details of 
common envelope evolution are still rather uncertain (see Sandquist 
et al. 1998 for a review), but in those systems where the orbital 
energy of the binary is not sufficient to eject the hydrogen 
envelope, the helium core of the giant and the compact remnant 
will merge.  Although it is difficult to separate the numerics 
from the physics in modeling the common envelope phase, modeling 
the merger of the helium core (which is several orders of magnitude 
smaller than the hydrogen giant) with a compact remnant is much 
more tractable.  

We model the merger using the three-dimensional smooth particle
hydrodynamics (SPH) code developed 
by Davies, Benz, \& Hills (1991).  To model the physics of the merging
process more accurately, the code has been improved (Fryer et al. 1999b)
to include an equation of state by Blinnikov, Dunina-Barkovskaya, 
\& Nadyozhin (1996). To follow nuclear reactions, we include a nuclear 
reaction network for temperatures above $4 \times 10^8\,\rm{K}$ 
(Timmes, Hoffman, \& Woosley 2000).  
The structure of the helium core was taken from 
1-dimensional helium star models by Heger (1999) which were mapped into
3-dimension and then allowed to settle into a stable configuration. For
all of our helium cores, the radius settled to within 10\% of the 
1-dimensional model.  

\begin{figure} 
\epsscale{1.0}
\plotone{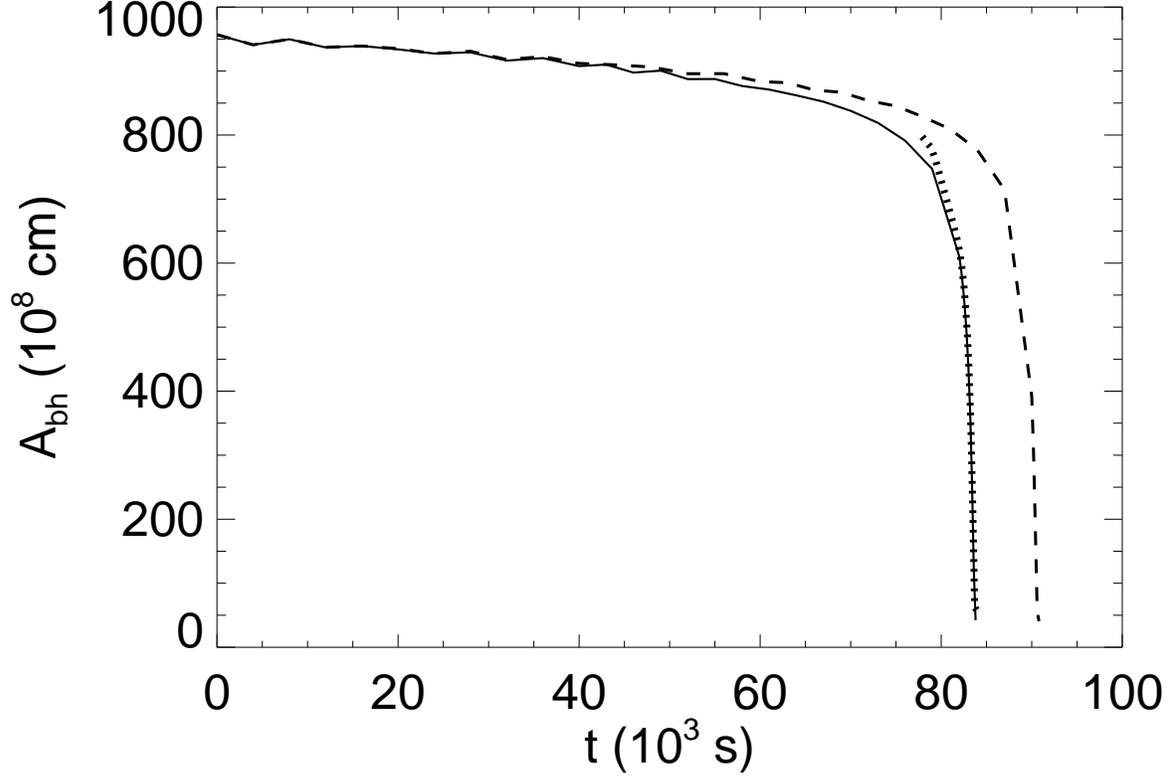}
\caption{The distance between the compact remnant and the center of the 
mass vs. time. The solid line show results with large initial separation for 
our first accretion recipe, which simply removes nearby particles around 
the compact remnant. The dotted line show results with small initial separation
for the first accretion recipe. The beginning of dotted line is at 80000\,s. 
The dotted line has been shifted along the time axis to compare the effects 
of our initial separation on the late-time inspiral. Note that the inspiral 
in the core is nearly identical for both initial separations.  
The dashed line show results with large initial separation
for our second accretion recipe, which gradually removes the mass from 
particles as they accrete onto the nearby compact remnant. \label{fig:AVSt}}
\end{figure}

Once the helium core has stabilized, we add the compact remnant (a point 
gravity source) and place the binary in a co-rotating circular orbit with 
a sufficiently small separation so that Roche-lobe overflow will occur.  
At this point, mass transfer from the helium core onto the compact remnant 
will tighten the orbit very rapidly and cause the compact remnant to merge 
with the helium core.  The orbital separation at which Roche-lobe
overflow commences is given by (Eggleton 1983): 
\begin{equation} 
A_0=R_{\he}\frac{0.6q^{2/3}+\ln (1+q^{1/3})}{0.49q^{2/3}},
\end{equation}
where $R_{\he}$ is the radius of the helium core,
$q=M_{\he}/M_{\BH}$
is ratio of the mass of the helium core $M_{\he}$ to the mass of the 
compact remnant $M_{\BH}$.  For our simulations, we are concerned 
primarily with the effects of the inspiral on the core of the helium 
star and these results are insensitive to our initial separation 
of the binary (that is, we can decrease the separation by $\sim$20\% 
and our results do not change).  For most of our simulations, we use 
initial separations smaller than $A_0$ so that the objects merge 
without detailed resolution of the outer layers of the helium star.
In addition, it is likely that the hydrogen common envelope phase 
would cause the compact remnant to spiral in below the $A_0$, so 
using the smaller separations may even be closer to the initial 
conditions produced by nature.  In any event, our simulations show that
two models with different initial separation have similar results 
(Fig.~\ref{fig:AVSt}).

Aside from initial separations, we must also be cautious with two other
numerical uncertainties.  First and foremost is the accretion onto the
compact remnant.  As the compact remnant merges with the helium star, 
the accretion rate onto the compact remnant is well approximated by the
Bondi-Hoyle accretion rate (Chevalier 1993; Brown 1995; Fryer, Benz,
\& Herant 1996; Bethe \& Brown 1998).  One might argue that if the 
compact remnant is a neutron star, photon pressure can slow the 
accretion.  However, the high densities in the helium core trap 
the photons and the time required for the photons produced 
at the hot surface of the neutron star to diffuse even 10\,km from 
the surface would take longer than our entire simulation time 
(see Fryer et al. 1996 for diffusion times).
However, the neutron star surface is hot enough to produce 
copious neutrinos which can cool the accreting material in 
less than 0.1\,s and this cooling (which however does not 
provide pressure support as photons would) 
allows the infalling material to accrete onto the neutron star 
(see Fryer et al. 1996).  Chevalier (1989) argued that even 
if the compact remnant had dipole magnetic fields in excess of 
$10^{15}$\,G, these fields would be 
buried by the high accretion flows that 
occur in He-mergers.  The more conservative analysis 
of Fryer et al. (1996) argues that the compact remnant's magnetic 
field would have to exceed $10^{13}$\,G to affect the accretion.
Such highly magnetized neutron stars are probably not common.  
Hence, for neutron star and black hole remnants alike, the inspiralling 
compact remnant will accrete at the Bondi-Hoyle
rate.  However, actually modelling this accretion can be quite difficult.

We try two methods to simulate the hypercritical accretion. In the
first accretion recipe, we simply remove particles that fall within 
5-10$\times 10^8\,\cm$ of the compact remnant and add their mass and
momentum to the compact remnant.  Although this is much larger than 
the Schwarzschild radius of the compact remnant, it is the smallest 
scale we can achieve given our resolution (1-3 $\times 10^9\,\cm$).

\begin{figure} 
\epsscale{1.0}
\plotone{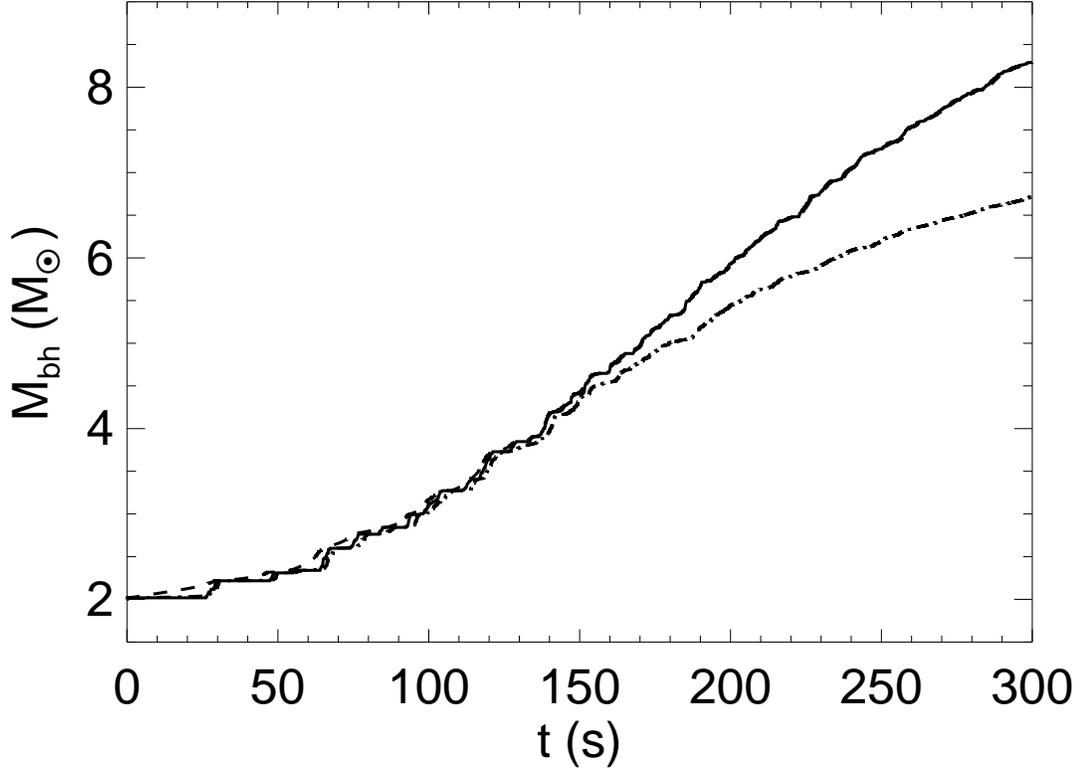}
\caption{The mass of the compact remnant vs. time for a central object 
accreting an $8\Msun$ helium star. The solid line shows results 
for the first recipe, which simply removes nearby particles around the
compact remnant, the dashed line shows results for the second recipe,
which gradually removes the mass from particles as they accrete onto
the nearby compact remnant.  The accretion rate for this second recipe
is determined by the free-fall timescale.  As the angular momentum of
the accreting material increases, the viscous timescale necessary to
remove this angular momentum can dominate the accretion timescale.  
The dash dot line shows results for a variation of the second recipe,
where we assume that the accretion timescale is 10 times the free-fall
timescale (assuming that the viscous timescale is much larger than the 
free-fall timescale).  Because the particles do not accrete as 
quickly, they build up around the compact remnant, creating higher
densities and, hence, shorter accretion timescales and the total
accretion rate only changes 10-20\% even though we decreased the
timescale by a factor of 10!  \label{fig:mVSt}}
\end{figure}

In the second accretion method, when the distance between a particle
and the compact remnant, $r_{\rm p}$, is less than the scale length
$h$ of that particle, the mass of that particle $M_{\rm p}$ is
accreted onto the compact remnant with a rate given by $\dot{M}_{\rm
p} = M_p/t_{\rm ff}$, here $t_{\rm ff}$, the free-fall timescale is
$\sqrt{2 r_{\rm p}^3/G M_{\BH}}$, where $G$ is the gravitational
constant and $M_{\BH}$ is the compact remnant mass. In this method, if
its distance from the black hole is less than $h$, the particle's
motion is dominated by gravitational force of the compact remnant and
it loses mass on the free-fall timescale.  We reduce the particle's
mass gradually as it accretes onto the compact remnant.  The second
accretion method yields a slightly longer inspiral time (by 10-20\%:
see Fig.~\ref{fig:AVSt}) which, in turn, yields accretion masses which
differ by roughly the same amount (see \S~\ref{sec:res}).  To test the
differences in the accretion {\it rate}, we ran a test model where we
placed a $2\,\Msun$ compact remnant in the center of an $\sim
8\,\Msun$ static helium core and measured the accretion as a function
of time (Fig.~\ref{fig:mVSt}).  Here we see that the accretion rates
for both our accretion mechanisms are nearly identical.

One might argue that the accretion timescale is not the free-fall
timescale because angular momentum can slow the accretion.  However,
during the inspiral stage of the merger of a helium and a compact
remnant, the angular momentum of particles accreted by the compact
remnant is small.  For the duration of our simulation, the accretion
timescale is dominated by the free-fall timescale, not by the viscous
timescale of the accretion disk.  However, to test the differences
caused by an increased accretion timescale due to the additional
effects of a viscous timescale, we modified the second accretion
method by assuming the accretion timescale was 10 times the free-fall
timescale and ran our test model with this increased timescale
(Fig.~\ref{fig:mVSt}).  The accretion rate for this factor of 10 range
in timescales is nearly identical for the first 150\,s and does not
differ by more than 20\% for the entire simulation.  

This is not a surprising result.  By increasing the accretion
timescale, we initially lower the accretion rate onto the black hole,
but this does not lower the rate at which material piles up around the
black hole (which is what our hydrodynamic simulation is modeling).
When we raise the accretion timescale, more material piles up around
the black hole, and the density of this material increases, causing
the accretion rate to rise.  Hence an equilibrium is achieved.  The
accretion rate onto the black hole is more sensitive to the rate at
which the material piles up around the black hole than upon the
accretion timescale of any individual particle.  This is nearly
identical to the effect seen in accretion disk simulations where the
accretion rate from the innermost point of the disk onto a black hole
is governed more by the rate at which material is put into the disk
rather than the actual value of the artificial viscosity (Popham,
Woosley, \& Fryer 1999; MacFadyen \& Woosley 1999).  Because the
accretion rate onto the black hole is relatively insensitive to our
accretion recipe and because the first recipe, which simply removes
nearby particles around the black hole (or neutron star), runs faster
than other recipes, we use the first recipe in all other simulations.

\begin{figure} 
\epsscale{1.0}
\plotone {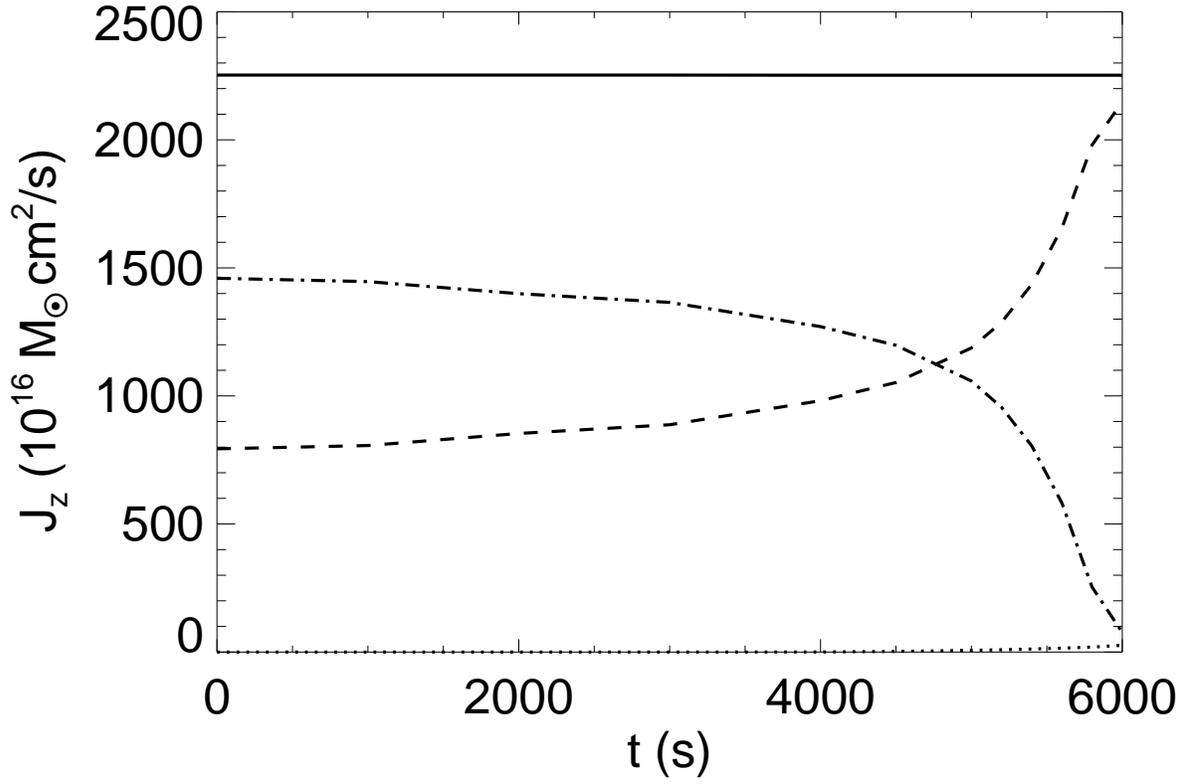}
\caption{The angular momentum vs. time. The solid line denotes total 
angular momentum, the dashed line denotes
angular momentum of the helium core, the dash dot lines denotes angular 
momentum of the compact remnant, and the dotted line denotes the angular 
momentum which is not accreted in the disk and is discarded in our 
simulations. \label{fig:jVSt}}
\end{figure}

In both methods, we also must add the angular momentum of the particles to 
the compact remnant.  Unfortunately, we do not resolve the accretion disk 
which must form around the compact remnant and so cannot model the transport 
of angular momentum out of this disk as it accretes.  Instead, we constrain 
the amount of angular momentum accreted to the angular momentum of the 
last stable orbit of the compact remnant.  We assume the rest of the angular 
momentum is transported out of the system.  Because our simulation time
step is much longer than the orbital period of the last stable orbit, this 
is a reasonable assumption.  This angular momentum may feedback into the 
star, and we do not include this.  But the angular momentum 
that we discard in our simulation is less than 1\% of the total angular 
momentum in the system (Fig.~\ref{fig:jVSt}), so we do not underestimate 
the angular momentum in the helium core by more than 1\% at the end 
of our simulation.

Figure~\ref{fig:jVSt} also shows how well the total angular momentum 
in our simulations is conserved.  For the merger of a $2\,\Msun$ compact
remnant and an $\sim 8\,\Msun$ helium star, the summation of total angular 
momentum of particles, total angular momentum of the compact remnant 
(including the estimated spin angular momentum) and the angular 
momentum that we discard during accretion is conserved 
to one part in ten thousand.  During the course of the simulation, 
nearly all of the orbital angular momentum of the compact remnant is 
converted into spin angular momentum of the helium core, and yet the 
total angular momentum is conserved!

The other major numerical uncertainty is the artificial viscosity used
in the SPH.  We have varied the artificial viscosity by a factor of 5 
to determine our dependence upon the numerical viscosity.  We will 
discuss these results in \S~\ref{sec:res}, but the quick answer is 
that simulations using this wide range of artificial viscosity obtain 
nearly the same results and our estimates of GRB energies and disk 
structures do not depend on the artificial viscosity.

\section{Results} \label{sec:res}

We have run a series of models testing the initial helium 
star masses, initial separations, and the range of uncertainties 
in the numerical methods (accretion algorithm, artificial viscosity, 
resolution, etc.).  B\"ottcher \& Fryer (2000) found that the helium 
star masses in He-merger progenitors cover a wide range of masses:  
14\% have masses less than $7\Msun$ and 22\% have masses above $12\Msun$ 
with the bulk (64\%) lying within $7-12\Msun$.  By varying the mass of the 
helium star, we can study the range of burst properties from 
He-mergers.  The entire set of simulations is summarized in 
Table~\ref{tab:par}. The number in the model designation denotes the
helium core mass in units of solar mass. 

We break up the formation of a He-merger GRB engine into two 
stages:  the inspiral of the compact remnant (which we model with 
our SPH code) and a post-processing analysis of the accretion onto 
the compact remnant once it reaches the core of the helium star.
In our simulations, we follow the compact remnant as it spirals into the 
helium star and accretes mass from the helium star while spinning up 
the star.  When the compact remnant reaches the center of 
the helium core, our simulations provide the mass and spin of the 
compact remnant as well as the structure and angular momentum 
distribution of the helium star.

The results from our simulations allow us to estimate the accretion 
of the rotating helium core around the now central black hole.  
We know that this material will accrete rapidly (especially along 
the spin axis of the core), but the material along the equator has 
too much angular momentum to accrete immediately and will first form a disk
around the compact remnant.  It is this disk accretion that can drive
a GRB along the relatively clear spin axis.  The mass of the compact 
remnant and the spin of the accreting material determine the viability
of He-mergers as a GRB engine.  

\subsection{Inspiral} \label{subsec:ins}

We presume that the compact remnant has already spiralled through 
the hydrogen envelope of a massive giant star (via a common envelope 
phase) and begin our simulations as the helium core contacts the 
compact remnant in Roche lobe overflow.  For two of our simulations 
(8D and 8E) we placed the compact remnant in a circular orbit with 
an orbital separation just low enough that the helium star overfilled 
its Roche lobe and accreted onto the compact remnant.  Even if orbital
angular momentum were conserved, this mass transfer (from a more 
massive object onto a less massive object) would tighten the orbits.  
But, in fact, a good deal of angular momentum is lost as material 
is ejected from the system.  In these simulations, we have implicitly 
assumed that when Roche-lobe overflow occurs, this mass transfer
dominates the orbital evolution of the binary.  

The viscosity caused by the hydrogen envelope may dominate the inspiral 
even after the helium core has overfilled its Roche lobe and is 
accreting onto the compact remnant (although we know that when 
the compact remnant actually enters the helium core, the effects of 
the helium core must dominate the inspiral).  Fortunately, most of 
the accretion occurs after the compact remnant has spiralled deep 
into the helium star and the merger evolution below a separation 
of $\sim 5 \times 10^{10}$\,cm is not affected by our initial 
separation (Fig.~\ref{fig:AVSt}).  Thus, for our $16\,\Msun$ and 
$4\,\Msun$ helium cores, we only model the inner $\sim 70\%$ of the 
core and start our simulations with smaller separations. This allows 
us to better resolve the core with a given number of particles.

\begin{figure}
\epsscale{1.0}
\plotone{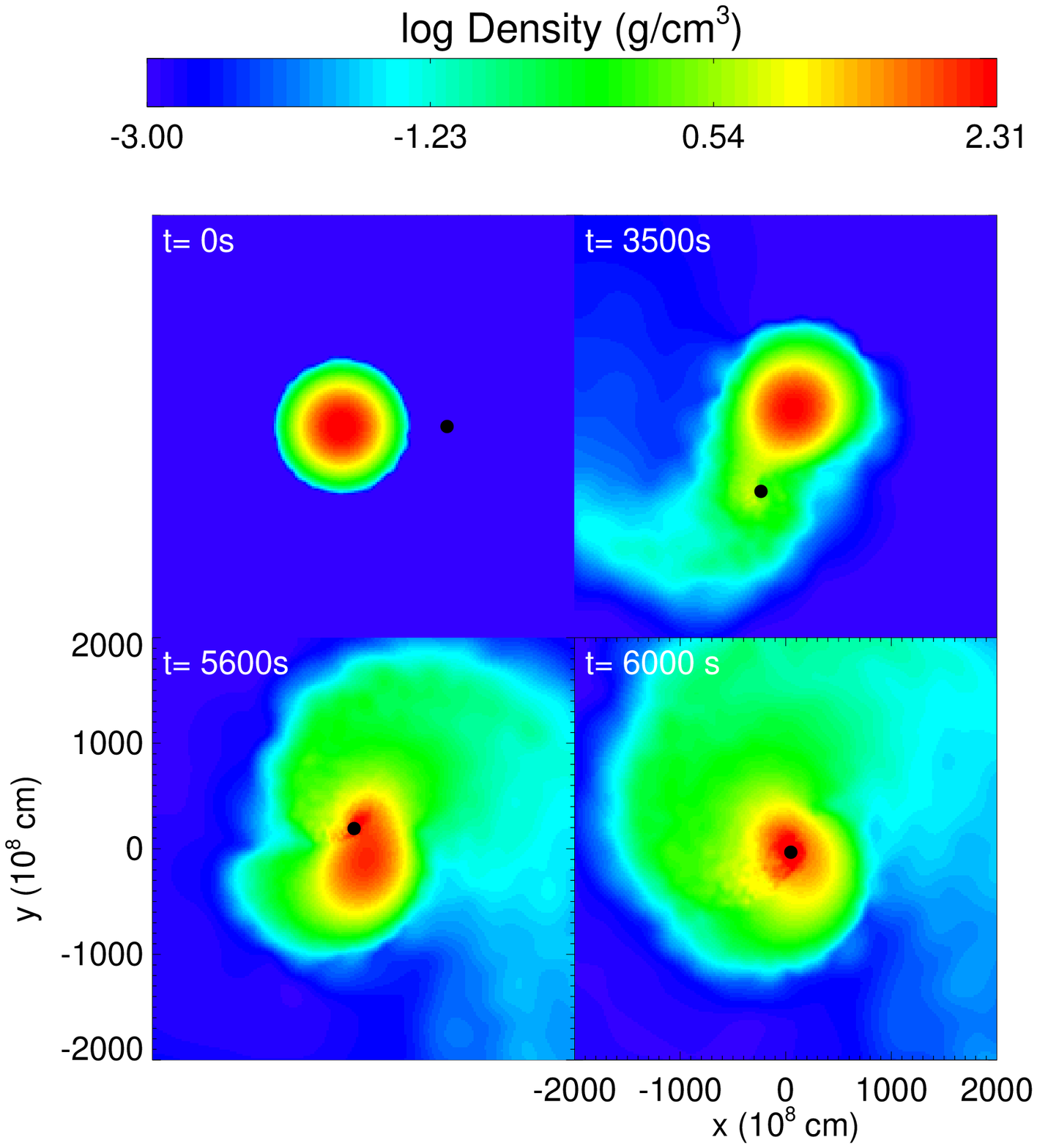}
\caption{Merger of a $2\,\Msun $ compact remnant with a $8\,\Msun $ helium
core. \label{fig:h8}}
\end{figure}

\begin{figure}
\epsscale{1.0}
\plotone{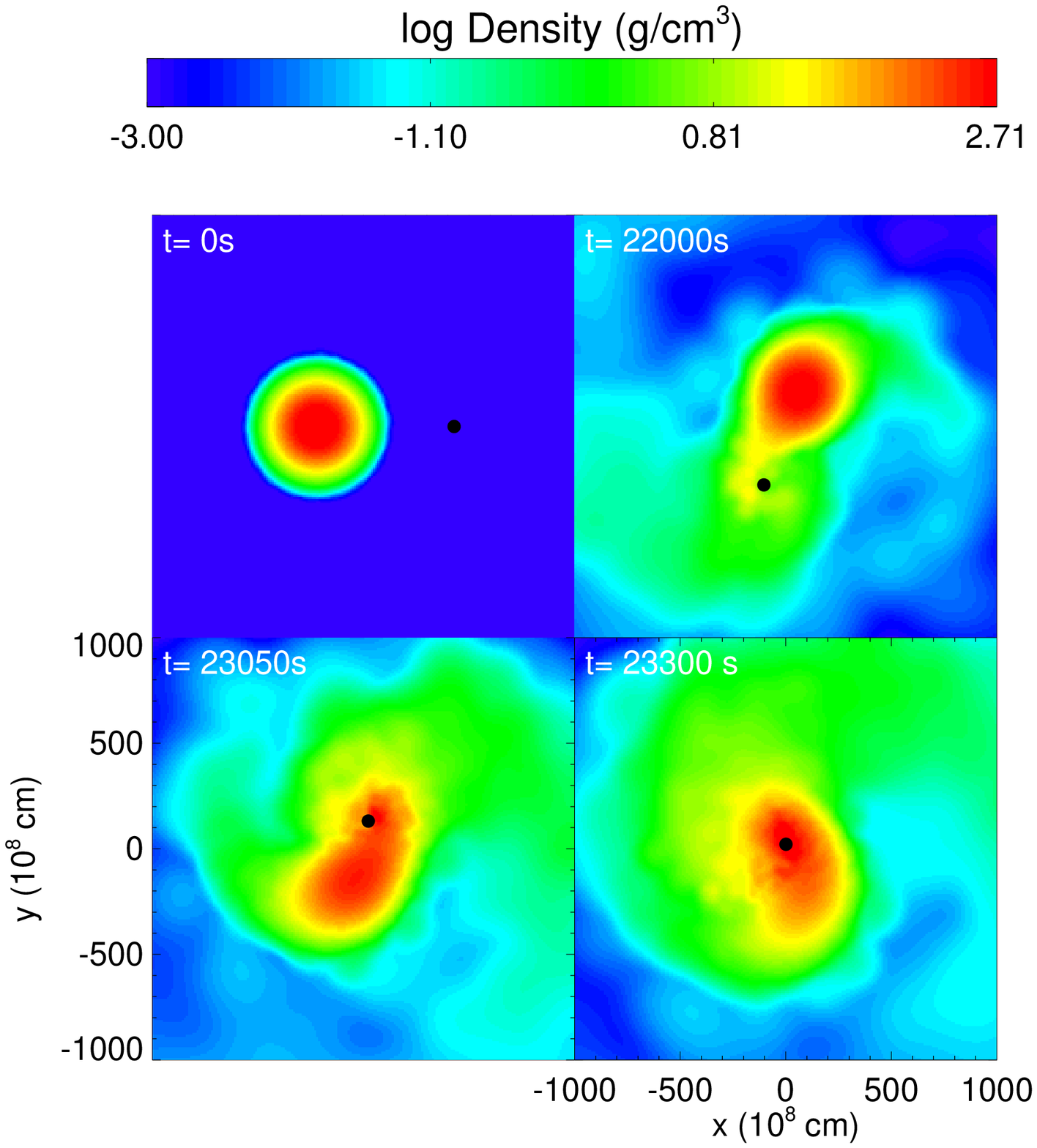}
\caption{Merger of a $2\,\Msun $ compact remnant with a $4\,\Msun $ helium
core. \label{fig:h4}}
\end{figure}

\begin{figure}
\epsscale{1.0}
\plotone{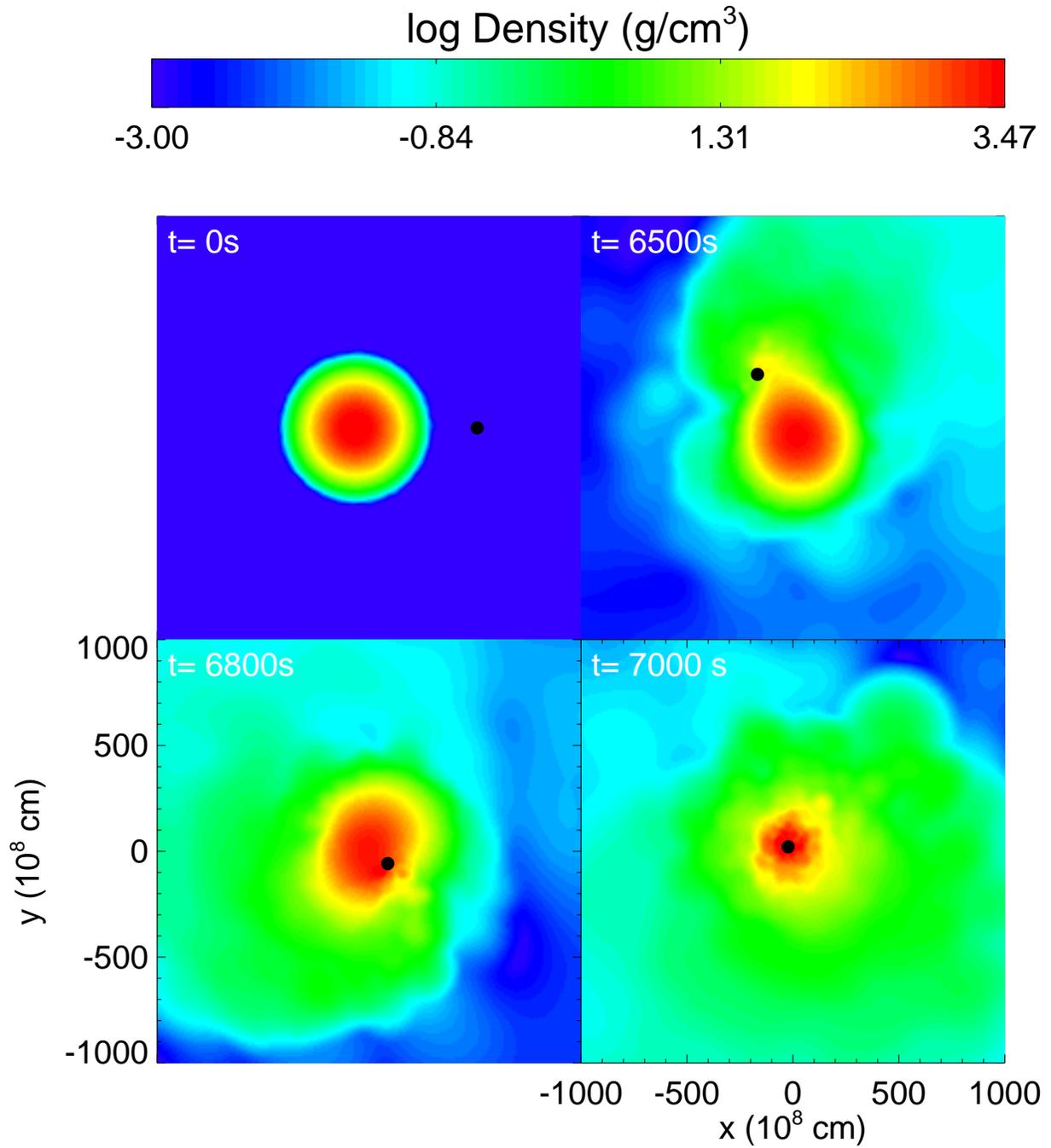}
\caption{Merger of a $2\,\Msun $ compact remnant with a $16\,\Msun $ helium
core. \label{fig:h16}}
\end{figure}

Some snapshots of the inspiral process for models 8B, 4A and 16A are 
shown, respectively, 
in Figures~\ref{fig:h8}, \ref{fig:h4} and \ref{fig:h16}.  Some material
is ejected, carrying away angular momentum and some mass.  In 
general, this ejecta does not play an important role on the GRB 
explosion itself because it is likely that the GRB explosion is 
beamed along the axis of rotation and the ejecta lies along 
the equator.  But this ejecta may effect the environment around 
the GRB and may have some observational features (e.g. iron lines -
B\"ottcher \& Fryer 2000).

\begin{figure}
\epsscale{0.8}
\plotone{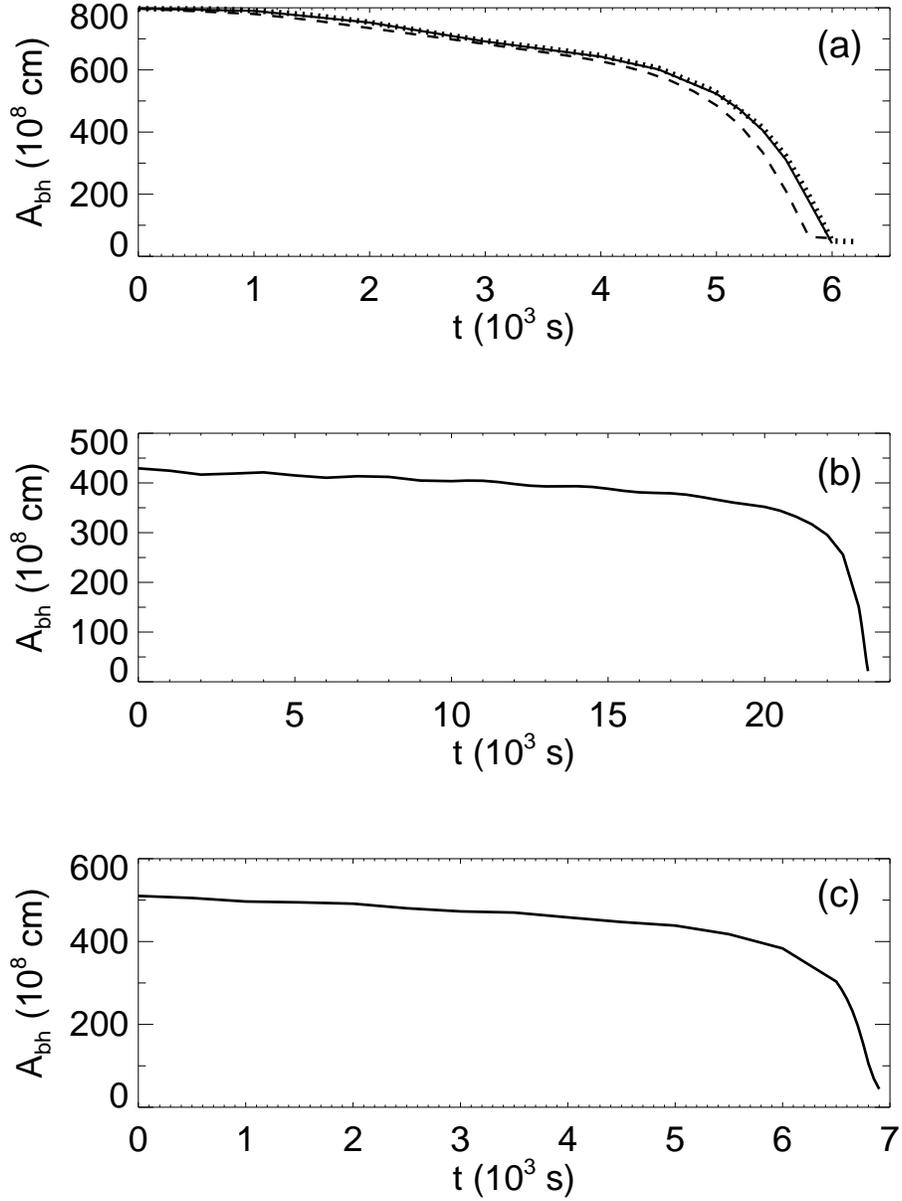}
\caption{The distance between the compact remnant and the center of mass 
is plotted as function of time for different models: (a) model 8A (solid
line), 8B (dotted line), and 8C (dashed line); (b) model 4C; 
(c) model 16A.  The compact remnant inspirals from half its initial
separation to the center of the helium star in less than $\sim$ 500\,s.
\label{fig:AVStall}}
\end{figure}

Although the binary can survive many orbital periods in the Roche-lobe
mass-transfer phase, as soon as the compact remnant begins to merge 
with the helium core, the orbital separation evolves very rapidly.  
The compact remnant inspirals from half its initial separation to 
the center of the helium star in less than $\sim$500\,s 
(Fig.~\ref{fig:AVStall}).  It is during this latter evolution that most 
of the accretion onto the compact remnant occurs and the conditions 
for a GRB engine are produced.  Changing the artificial viscosity by 
a factor of 5 changes this inspiral time by less than 10\%.  Also, 
results of high resolution model 8B are similar to those of low 
resolution model 8A (Fig.~\ref{fig:AVStall}).

In the initial stages of the inspiral, the accretion rate onto the 
compact remnant is relatively low (less than $10^{-4}\,\Mss $) and 
the compact remnant gains very little mass.  But during the rapid 
inspiral phase, the accretion rate increases dramatically and nearly 
all of the accretion occurs in the last few hundred seconds.
We stop our simulations when the compact remnant reaches 
the ``center'' of the helium star.  

After the inspiral stage, the effect of angular momentum which can
slow the accretion gradually becomes important. However, at the end of
our simulations, the effect of angular momentum is still small.  
For example, at the end of the simulation for model 8B, the
viscous timescale of most of the matter (93\% of particles) accreting 
onto the black hole is less than the free-fall timescale (we estimate
the viscous timescale by: $t_{\rm visc}=(\alpha \omega)^{-1}$
where $\alpha=0.1$ is the disk viscosity and $\omega$ is the orbital 
angular velocity).  However, the effect of angular momentum is increasing
rapidly, and if we want to model the accretion beyond the inspiral phase, 
we would need to modify our accretion recipe in addition to increasing 
our mass resolution.

\begin{figure}
\epsscale{1.0}
\plotone{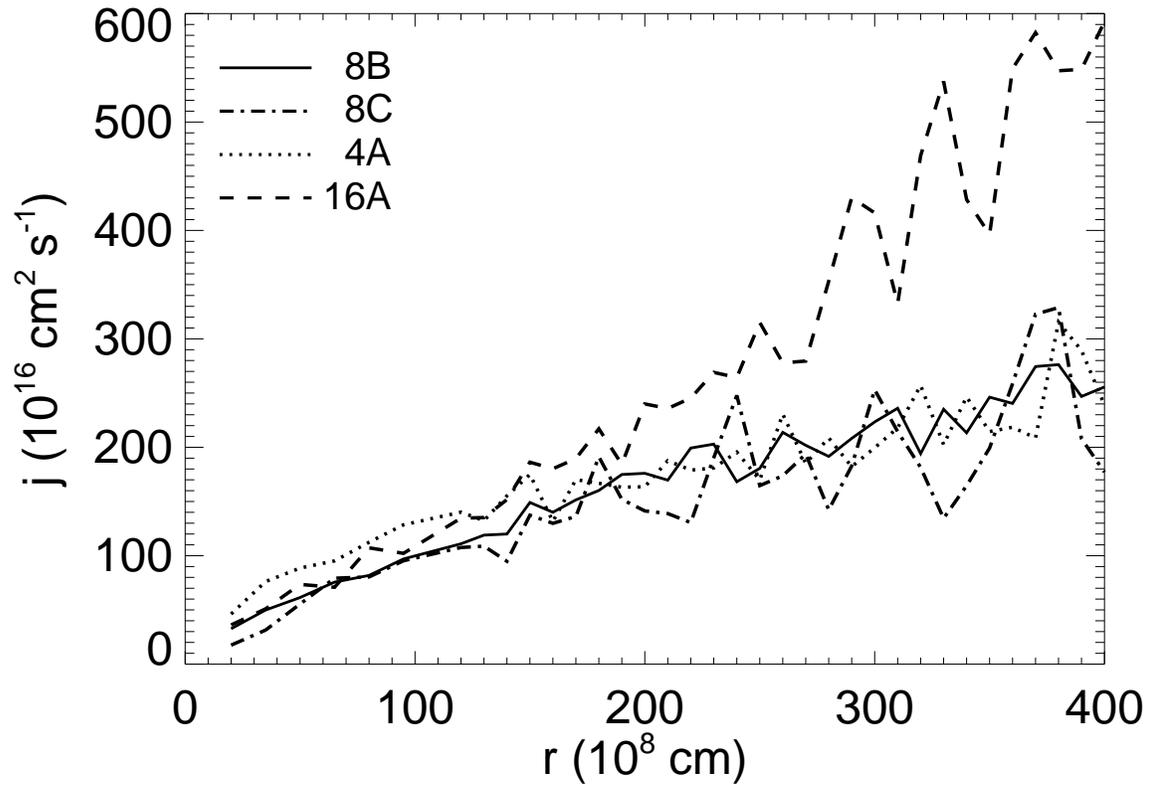}
\caption{Spin angular momentum of the helium core at the time when the
compact remnant reaches the center of the core for model 8B (solid line),
8C (dash dot line), 4A (dotted line) and 16A (dashed line). 
\label{fig:jr}}
\end{figure}

The mass of the compact remnant at the end of our simulations are
given in Table~\ref{tab:erg}.  At this stage, nearly all of the
orbital angular momentum of the system has been converted to spin
angular momentum in the helium star (Fig.~\ref{fig:jr}).  Some of the
orbital angular momentum is converted into spin of the compact remnant
(Table~\ref{tab:erg}).  This merged system is now very similar to that
of a collapsar (a rotating helium star collapsing onto a central black
hole).  These conditions are believed to drive GRB explosions, but to
estimate the explosion energy, we must calculate the later accretion.

\subsection{Accretion Rate} \label{subsec:mdot}

Our simulations show that He-mergers will indeed produce rapidly 
rotating helium stars accreting onto a central compact remnant in 
much the same manner as is found in the collapsar GRB engine.
However, the angular momentum in He-mergers (Fig.~\ref{fig:jr}) 
is roughly an order of magnitude greater than that of the collapsars 
studied by MacFadyen \& Woosley (1999). With such high angular
momentum, following the continued accretion requires an understanding 
of disk accretion.  Even if we assumed an artificial angular 
momentum transport algorithm, we would need much more resolution 
than we have in our current simulations to follow disk formation and the
accretion onto the compact remnant.

However, we can analytically estimate the accretion rate onto 
the compact remnant by estimating the accretion time for each particle.  
Such a rough calculation can be obtained by assuming that pressure 
support is negligible and gravitational and centrifugal forces
dominate the motion of the particles (for most of the star, this is true).

We take the end of our SPH simulations as the initial conditions of 
our analytic estimate.  Without pressure support, each particle
accelerates to nearly the free-fall velocity until centrifugal support
slows its collapse and it accretes onto a disk.  We can derive the 
radius at which this occurs by knowing the enclosed mass ($M_{\rm
Enclosed}$) and the angular momentum ($j$), and then equating 
forces:  $r_{\rm disk}=j^2/GM_{\rm Enclosed}$.  We estimate the timescale for 
the particle to accrete onto the disk is simply the free-fall timescale:
\begin{equation}
t_{\rm ff}=\sqrt{\frac{2 r_{\rm initial}^{3}}{G M_{\rm enclosed}}}
         - \sqrt{\frac{2 r_{\rm disk}^{3}}{G M_{\rm enclosed}}},
\end{equation}
where $r_{\rm initial}$ is the initial position of the particle.  
For matter along the pole, the angular momentum may not be sufficient 
to halt the collapse, and these particles accrete directly onto the 
compact remnant at roughly the free-fall time. 

But the bulk of the matter will accrete through a disk, and we must 
then estimate the disk accretion timescale.  For each particle in the 
disk, we can calculate the viscous timescale:
\begin{equation}
t_{\rm visc}=(\alpha \omega)^{-1},
\end{equation}
where $\alpha$ is the disk viscosity parameter 
(which we take to be $0.1$), and $\omega$ is the orbit angular velocity. 
The total accretion timescale of each particle is just the sum of its 
free-fall and viscous components ($t_{\rm acc}=t_{\rm ff}+t_{\rm
visc}$).  We can 
then easily estimate the accretion rate by calculating the number (and
total mass) of particles accreted at a given time ($t_0$) 
\begin{equation}
\dot{M}=\sum^i_{t_0-\Delta t < t^i_{\rm acc} < t_0+\Delta t}
\frac{m_i}{2 \Delta t},
\end{equation}
where $m_i$ and $t^i_{\rm acc}$ are the mass and accretion rate of 
particle $i$.  The accretion rates for our 3 
helium star masses are shown in Figures~\ref{fig:md08}, \ref{fig:md04}
and \ref{fig:md16}.  The accretion rate not only provides us 
with an estimate of the black hole mass as a function of time, but 
also the spin of the black hole.  There is so much angular momentum 
in the helium core that nearly all of the accreting material accretes 
through a disk, and the angular momentum of the material at its last 
stable orbit quickly spins up the black hole.

In Figures~\ref{fig:md08} - \ref{fig:md16}, the accretion rates 
seem to vary on short timescales, but such wide variations are likely
to be an artifact of our low resolution.  Higher resolution simulations 
will allow us to determine how much of the variation in the accretion 
rate is real (due to asymmetries caused in the inspiral) and how 
much is due to numerical resolution.

Throughout our calculation, we have tried to estimate an upper limit
to the accretion rate, assuming no thermal pressure and a high disk
viscosity.  Even so, except for the $16\,\Msun$ merger, the accretion
rates we obtain are much lower than those calculated in collapsars.
This is, in part, because our helium cores have much higher angular
momenta.  In addition, we modeled the merger of unevolved helium
cores.  If the common envelope phase occurred during Case C
mass-transfer, the merger would not have occurred until after helium
burning, and the more compact evolved core would yield higher initial
accretion rates.

\section{Gamma-Ray Bursts} \label{sec:GRB}

In this paper, we present the first hydrodynamical simulations of 
the formation of He-merger GRBs up to the formation of a central 
black hole around a rotating helium core.  With the conditions 
produced during the inspiral, it is clear that the compact remnant 
will collapse to a black hole (if it is not one already).  The 
densities around the central compact remnant are so high that photons 
are trapped in the material and radiation pressure does not 
halt accretion.  Even if the compact remnant is a neutron star, 
neutrino cooling will allow high accretion rates ($> 10^5 \Msyr$) 
which will quickly cause the neutron star to collapse to a 
black hole.  The accreting material will also spin up the compact 
remnant and He-mergers produce a population of rapidly rotating 
black holes.  With possible merger rates above $10^{-5} {\rm yr}^{-1}$
in the Galaxy, these black holes may make up a few percent of the 
total black holes 
formed in the galaxy.  However, they do not produce slowly accreting 
neutron stars (e.g. Thorne-Zytkow stars) and cannot be used 
to explain strange neutron star systems.

\begin{figure}
\epsscale{1.0}
\plotone{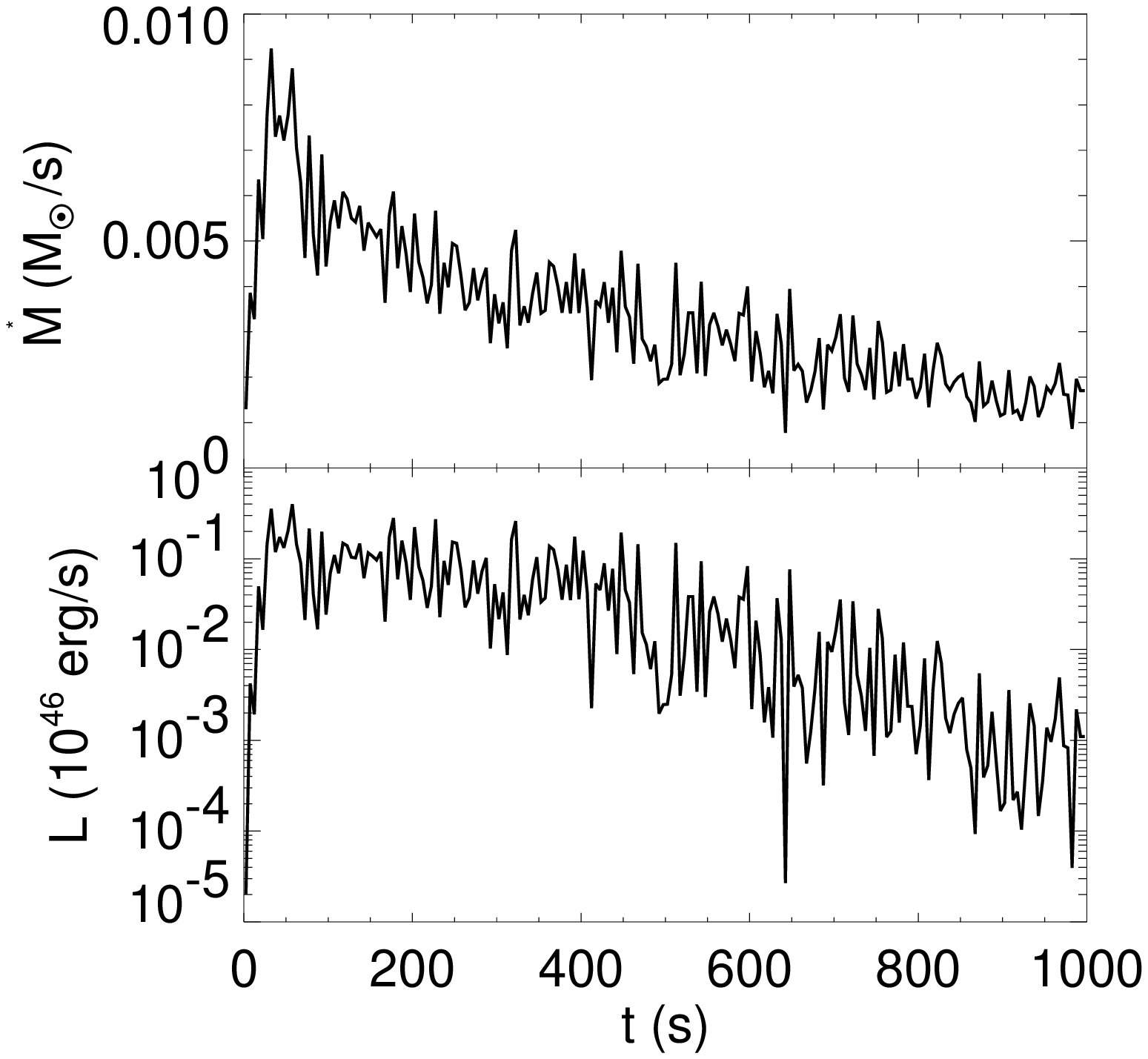}
\caption{Estimates of accretion rate and energy generation rate from
neutrino annihilation for model 8B. \label{fig:md08}}
\end{figure}

\begin{figure}
\epsscale{1.0}
\plotone{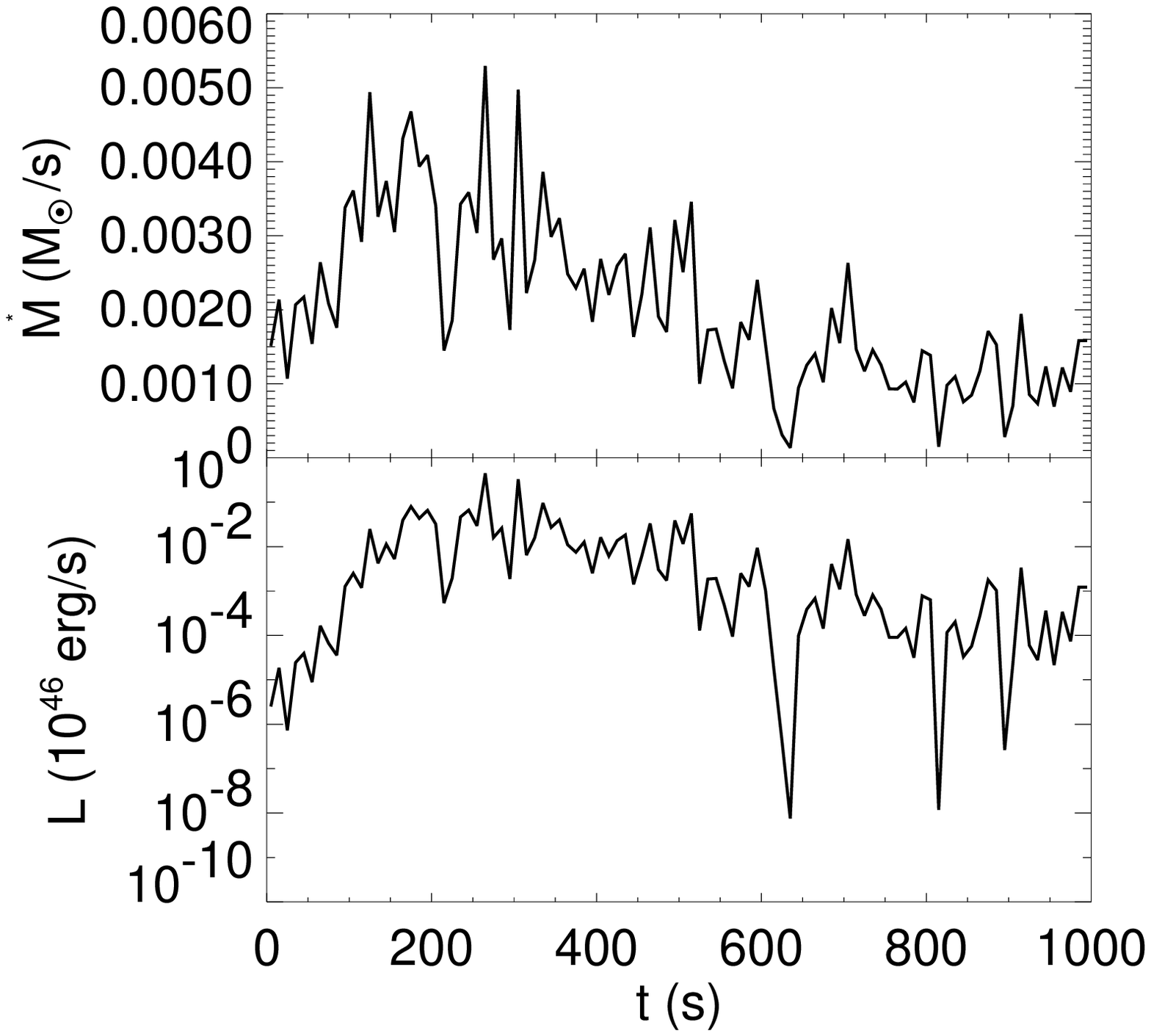}
\caption{Estimates of accretion rate and energy generation rate from
neutrino annihilation for model 4A. \label{fig:md04}}
\end{figure}

\begin{figure}
\epsscale{1.0}
\plotone{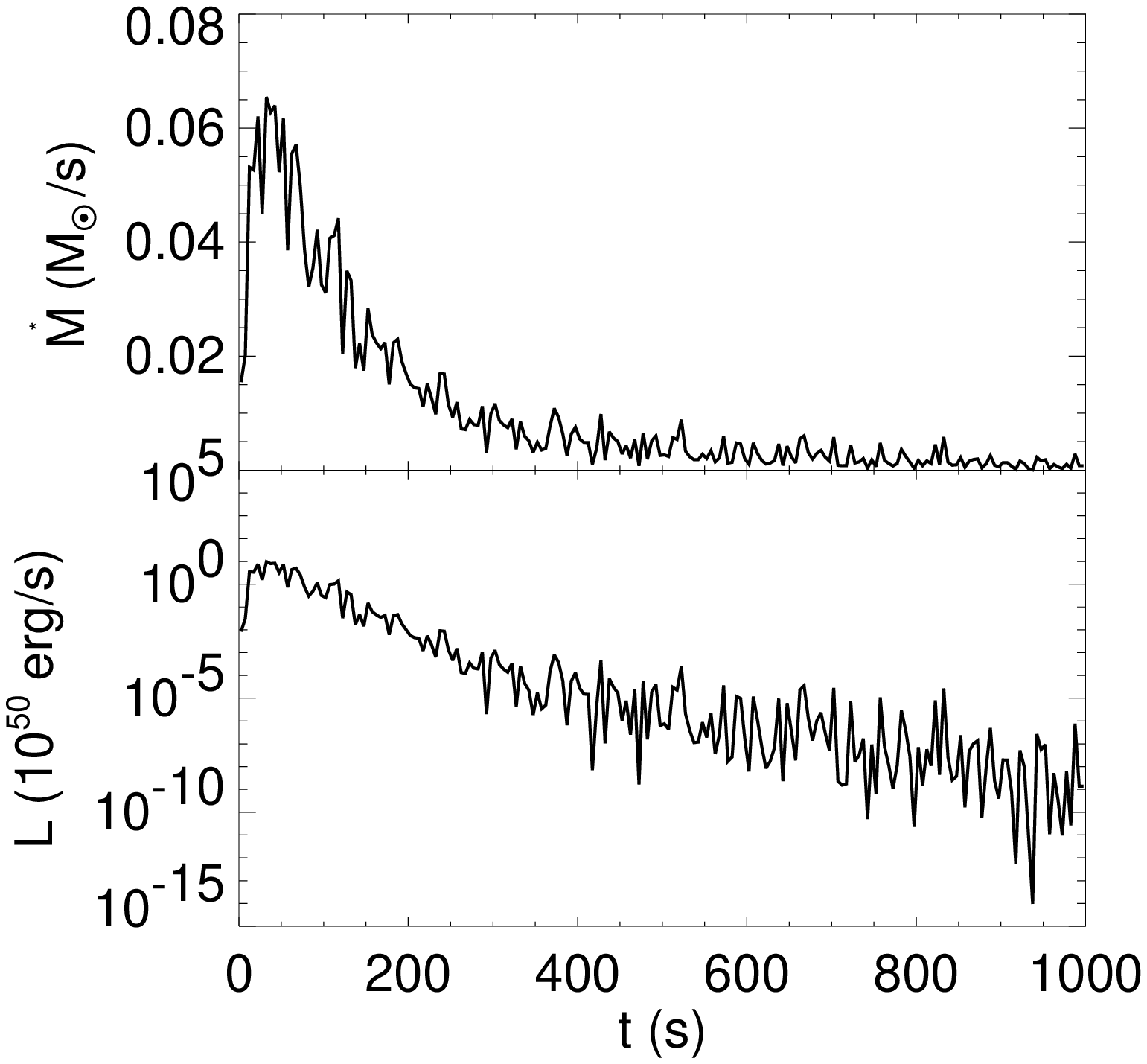}
\caption{Estimates of accretion rate and energy generation rate from
neutrino annihilation for model 16A. \label{fig:md16}}
\end{figure}

But can these mergers produce gamma-ray bursts?  The viability of the 
He-merger GRB model depends upon the energy of the explosion 
produced during the disk accretion phase.  The two most-often 
discussed power sources for these explosions are:  neutrino 
annihilation or some magnetic field mechanism akin to 
the engine behind the jets of active galactic nuclei.  The neutrino 
annihilation mechanism takes advantage of the profuse neutrino
emission which carries off much of the binding energy of the accreting
material.  Popham et al. (1999) calculated 
the energy emitted in neutrinos for hyper-accreting 
black holes. The subsequent energy 
generation rate from neutrino annihilation can be fitted by 
(Fryer et al. 1999b):
\begin{equation}
\log L_{\nu,\bar{\nu}}\,\es \approx 
43.6 + 4.89 \log \frac{\dot{M}}{0.01 \Mss} + 3.4 a,
\end{equation}
where $a\equiv J_{\BH} c/GM_{\BH}^2$ is the spin parameter of the black
hole.  Using this formula, and our derived accretion rates, black hole
spin evolution, etc., we can 
estimate the energy deposition rate ($L_{\nu,\bar{\nu}}$) for helium 
mergers of different initial helium star mass
(Fig.~\ref{fig:md08}-\ref{fig:md16}).  We stress that the variation in 
the energy deposition is almost certainly due to our low numerical
resolution, and it will require much higher resolution simulations which
can resolve any clumping structures to determine what fluctuations, if 
any, are produced in the accretion.  Integrating $L_{\nu,\bar{\nu}}$,
we get the estimates of the total explosion energy (Table~\ref{tab:erg}).
With beaming into 3\% of the sky, Model 16A produces 
an equivalent isotropic energy of $10^{54}$\,ergs!  
However, we should bear 
in mind that there are many uncertainties in these calculations:  
variations in the numerics (compare models 8A, 8B, 8C), simplified 
assumptions in our accretion rates, and the uncertainties in the neutrino 
annihilation deposition.  The errors in our calculations of the 
explosion energy may exceed 1 order of magnitude.  Because we chose 
high disk viscosities and assumed none of the energy produced would 
be dragged back into the accreting remnant, our estimates are likely 
to be upper limits to the burst energies.  Even so, except for 
He-mergers involving massive helium cores ($\sim 16 \Msun$), helium 
mergers powered by neutrino annihilation cannot produce the most energetic GRBs.
However, 22\% of He-mergers involve the merger of compact remnants and 
massive helium stars ($>12 \Msun$:  B\"ottcher \& Fryer 2000) and the 
rate of massive helium-star mergers may well be high enough to explain 
the observed rate of GRBs.

\begin{figure}
\epsscale{1.0}
\plotone{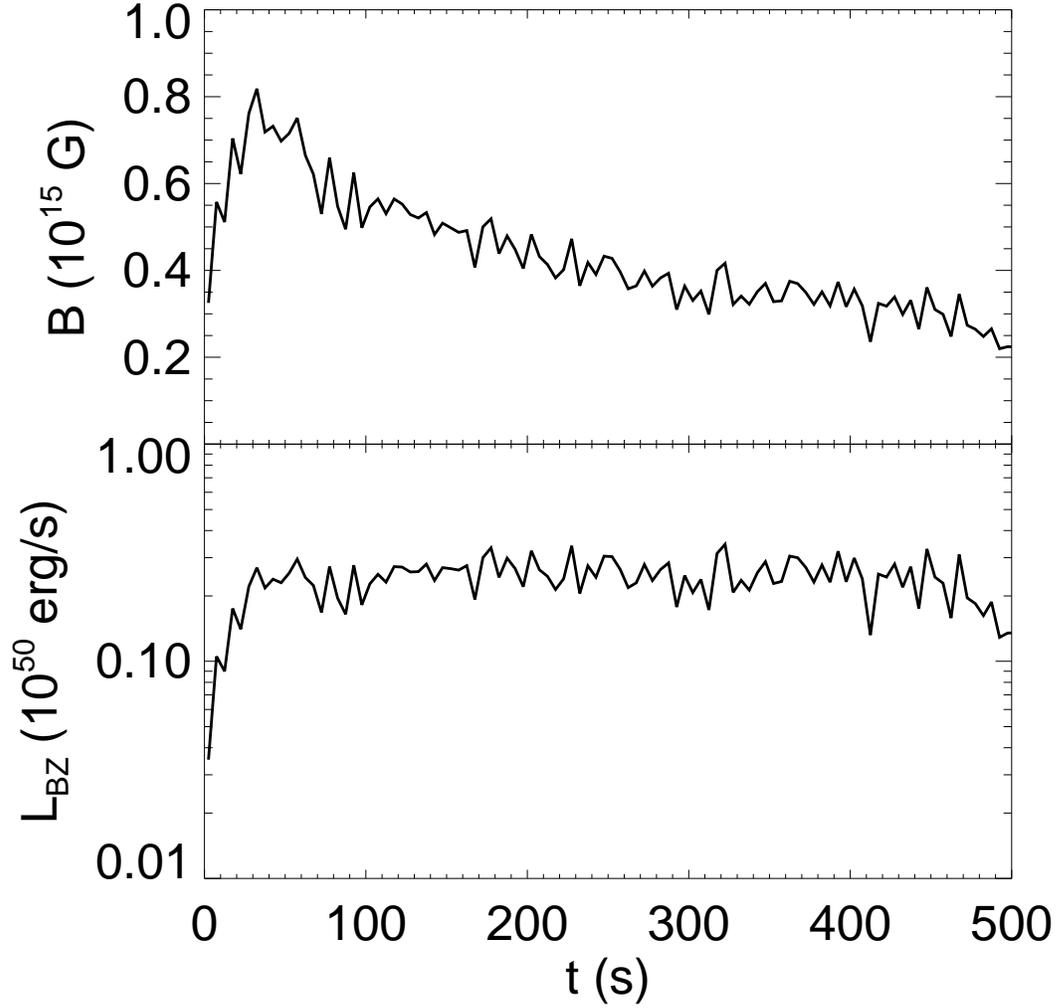}
\caption{Energy deposition rate vs. time from a GRB engine driven by 
magnetic fields for model 8B.  The energy does not drop off dramatically,
however, as soon as a burst is launched, it will disrupt the accretion
and shut the engine off.  These deposition rates assume the magnetic
energy density equals the kinetic energy density of the accreting 
material and is likely to be an overestimate. \label{fig:BZ08}}
\end{figure}

More promising for the low accretion rates in He-mergers, MHD processes
may extract the rotational energy of either the black hole or the accretion
disk and power the GRB (Blandford \& Znajek 1977; MacDonald et al 1986;
Paczy\'{n}ski 1991; Narayan, Paczy\'{n}ski, \& Piran 1992; Woosley 1993;
Hartmann \& Woosley 1995; Thompson 1996; Katz 1997; M\'{e}sz\'{a}ros 
\& Rees 1997; Popham et al. 1999). For example, the Blandford-Znajek 
luminosity is
\begin{equation}
L_{{\mathrm{BZ}}} \approx 10^{50} a^2 (\frac{M_{\BH}}{3\,\Msun})^2
(\frac{B}{10^{15}{\,\mathrm{Gauss}}})^2\,\es,
\end{equation}
where $a$ is the spin parameter of the black hole and $B$ is the magnetic
field strength in the disk.  We know the spin rate ($a$) and mass 
($M_{\BH}$) of the black hole and we can get an rough upper limit 
to the magnetic field strength by setting the magnetic energy 
density equal to the kinetic energy density of accreting disk. This yields:
\begin{equation}
B^2= 4 \pi \rho v^2 \approx \dot{M} c / r_{\rm g}^2, 
\end{equation}
where $r_{\rm g}\equiv 2 G M / c^2$.
Then we can estimate the maximum energy deposition rate for 
explosions driven by magnetic fields (Fig.~\ref{fig:BZ08}).  

When an explosion is produced, it can disrupt the disk and halt 
further accretion.  This will turn off the engine.  We assume the 
deposition continues for 200\,s (a long burst) and integrate the 
deposition rate to calculate maximum burst energies (Table~\ref{tab:erg}).  
Burst energies above $10^{51}$\,ergs are possible for all our
He-merger models.  These energies are most certainly overestimates, 
but lacking a working magnetic field mechanism, reliable calculations 
(with errors less than a few orders of magnitude) are impossible.  

However, it is clear that He-mergers {\it can} produce GRB explosions.  
Even with neutrino annihilation as the energy source, mergers of 
massive helium stars and compact remnants can explain long-duration 
bursts with inferred isotropic energies of above $10^{54}$\,ergs.
Because He-mergers are similar to collapsars (with slightly higher 
specific angular momenta), we believe they will have very similar 
characteristics to collapsars:  they are likely to be beamed, and will 
produce supernovae-like explosions (e.g. 1998bw).

\acknowledgements

This research has been supported by NASA (NAG5-2843, MIT SC A292701,
and NAG5-8128), the NSF (AST-97-31569), the US DOE ASCI Program 
(W-7405-ENG-48) and the Los Alamos Feynman Fellowship.  The authors 
gratefully appreciate stimulating conversations with Andrew MacFadyen
and Stan Woosley on the subject of gamma-ray bursts and Collapsars.  
We thank Stan Woosley and Tamara Rogers for helpful comments on the 
manuscript.

\clearpage

\begin{deluxetable}{ccccccc}
\tablecaption{Parameters of Models \label{tab:par}}
\tablewidth{0pt}
\tablehead{
\colhead{} &
\colhead{} &
\colhead{} &
\colhead{Initial} & 
\colhead{\# of} &
\colhead{Accretion} &
\colhead{Artificial} \\
\colhead{Model} &
\colhead{$M_{\rm Remnant}$} &
\colhead{$M_\he$} &
\colhead{Separation} &
\colhead{Particles} &
\colhead{Recipe} &
\colhead{Viscosity} \\
\colhead{} &
\colhead{($\Msun$)} &
\colhead{($\Msun$)} &
\colhead{($10^8\,\cm$)} &
\colhead{} &
\colhead{} &
\colhead{}
}
\startdata
 8A & 2 &  8 & 1000 & 10000 & First  & High \\
 8B & 2 &  8 & 1000 & 30000 & First  & High \\
 8C & 2 &  8 & 1000 & 10000 & First  & Low \\
 8D & 2 &  8 & 1200 & 10000 & First  & High \\
 8E & 2 &  8 & 1200 & 10000 & Second & High \\
 4A & 2 &  4 &  650 & 10000 & First  & High \\
16A & 2 & 16 &  575 & 10000 & First  & High 
\enddata
\end{deluxetable}

\clearpage

\begin{deluxetable}{cccccc}
\tablecaption{Energy \& Duration \label{tab:erg}}
\tablewidth{0pt}
\tablehead{
\colhead{Model} & \colhead{$M_i$\tablenotemark{a}} & 
\colhead{$a_i$\tablenotemark{a}} & 
\colhead{$E_{\nu,\bar{\nu}}$\tablenotemark{b}} &
\colhead{$T^{\rm burst}_{90}$\,\tablenotemark{c}} &
\colhead{$E_{{\mathrm{BZ}}}$\tablenotemark{d}} \\
\colhead{} & \colhead{$(\Msun)$} & \colhead{} &
\colhead{$(10^{49}\,\erg)$} &
\colhead{(s)} &
\colhead{$(10^{49}\,\erg)$}
}
\startdata
 8A & 2.25 & 0.14 & 0.51  & 370 & $<$630 \\
 8B & 2.87 & 0.60 & 0.049 & 510 & $<$480 \\
 8C & 2.76 & 0.43 & 0.090 & 520 & $<$500 \\
 4A & 2.15 & 0.21 & 0.018 & 460 & $<$220 \\
16A & 5.54 & 0.985 & 3700 & 65  & $<$5500 \\
\tablenotetext{a}{The mass and spin of the compact 
remnant when it reaches the center of the helium star.}
\tablenotetext{b}{These are maximum total energies emitted 
via neutrino annihilation.  We assume no energy is carried 
in along the poles.}
\tablenotetext{c}{The burst duration for neutrino annihilation 
driven explosions is the timescale in which 90\% of the total 
energy is emitted.}
\tablenotetext{d}{These are probably maximum total energies 
from a magnetically driven mechanism, assuming that the magnetic 
field energy density equals the kinetic energy of the accreting 
disk.}
\enddata
\end{deluxetable}

\clearpage

\clearpage

\clearpage

\clearpage

\clearpage

\clearpage

\clearpage

\clearpage

\clearpage

\clearpage

\clearpage

\clearpage

\end{document}